\documentclass[12pt]{article}

\ifx\pdfoutput\undefined
\usepackage[dvips,bookmarks]{hyperref}
\else
\usepackage{hyperref}
\fi
\hypersetup{colorlinks=false,bookmarksopen,bookmarksnumbered,citecolor=blue,
   pdfstartview=FitH}

\usepackage[dvips]{graphicx}
\usepackage{latexsym}

\usepackage{amssymb,amsfonts,amsmath}
\usepackage{graphicx} 
\usepackage{indentfirst}

 \usepackage{bbm}

\parskip=\medskipamount

\newcommand {\cD}{{\cal D}}

\newcommand {\cF}{{\cal F}}

\newcommand {\cL}{{\cal L}}
\newcommand {\cM}{{\cal M}}
\newcommand {\cN}{{\cal N}}

\newcommand {\cV}{{\cal V}}
\newcommand {\cW}{{\cal W}}

\def\a{\alpha}

\def\b{\beta}

\def\d{\delta}

\def\g{\gamma}

\def\m{\mu}

\def\o{\omega}

\def\q{\theta}

\def\s{\sigma}

\def\D{\Delta}
\def\F{\Phi}

\def\O{\Omega}

\def\Q{\Theta}

\def\rd{{\rm d}}
\def\ri{{\rm i}}

\newcommand{\ad}{{\dot{\alpha}}}                           
\newcommand{\bd}{{\dot{\beta}}}                            
\newcommand{\ve}{\varepsilon}                            
\newcommand{\cDB}{{\bar\cD}}                            

\newcommand{\pa}{\partial}                           
\newcommand{\hf}{\frac12}

\newcommand{\be}{\begin{equation}}
\newcommand{\ee}{\end{equation}}
\newcommand{\bea}{\begin{eqnarray}}
\newcommand{\eea}{\end{eqnarray}}
\newcommand{\non}{\nonumber}
\newcommand{\1}{{\underline{1}}}
\newcommand{\2}{{\underline{2}}}

\newcommand{\bm}[1]{\mbox{\boldmath$#1$}}

\def\double #1{#1{\hbox{\kern-2pt $#1$}}}

\newcommand{\gd}{{\dot\g}}
\newcommand{\dd}{{\dot\d}}

\newcommand{\ts}{{\tilde{\s}}}

\newcommand{\teb}{{\bar{\theta}}}

\begin{document}
\begin{titlepage}
\begin{flushright}
UMD-PP-09-048\\
September, 2009\\
\end{flushright}
\vspace{5mm}

\begin{center}
{\Large \bf Chiral supergravity actions and superforms}\\ 
\end{center}

\begin{center}

{\bf
S. J. Gates, Jr.,${}^{a}$\footnote{gatess@wam.umd.edu}
S. M. Kuzenko${}^{b}$\footnote{kuzenko@cyllene.uwa.edu.au}
and
G. Tartaglino-Mazzucchelli${}^{a}$\footnote{gtm@umd.edu}
} \\
\vspace{5mm}

\footnotesize{

${}^{a}${\it Center for String and Particle Theory,
Department of Physics, 
University of Maryland\\
College Park, MD 20742-4111, USA}
~\\

\vspace{5mm}

${}^{b}${\it School of Physics M013, The University of Western Australia\\
35 Stirling Highway, Crawley W.A. 6009, Australia}}  
~\\

\end{center}
\vspace{5mm}

\begin{abstract}
\baselineskip=14pt
The superform construction of supergravity actions, christened the ``ectoplasm method,''
is based on the use of a closed super $d$-form in the case of $d$ space-time dimensions.
In known examples, such  superforms are obtained by iteratively solving 
nontrivial cohomological problems. The latter usually makes this scheme no less
laborious than the normal coordinate method for  deriving component actions for matter-coupled 
supergravity. In this note we present an alternative procedure to generate  required superforms
in four space-time dimensions, which makes use of self-dual vector multiplets.
It provides the shortest derivation of chiral actions in two different theories:
(i) $\cN=1$ old minimal supergravity; and (ii) $\cN=2$ conformal supergravity. 
The $\cN=2$ superform construction is developed here for the first time. 
Although our consideration is restricted to the case of four dimensions, 
a generalization to higher dimensions is plausible.
\end{abstract}
\vspace{1cm}

\vfill
\end{titlepage}

\newpage
\renewcommand{\thefootnote}{\arabic{footnote}}
\setcounter{footnote}{0}

\section{Introduction}
\setcounter{equation}{0}
The power of  superspace approaches to supergravity theories 
in diverse dimensions consists in  the possibility
to write down the most general  locally supersymmetric actions formulated in terms 
of a few  dynamical variables with simple geometric origin. 
This generality does not come without  price to be paid.
The point is that, being trivial in principle, 
a reduction from the parental superfield action 
to its component counterpart requires some work that is
technically quite involved and challenging in many concrete cases.

For off-shell supergravity theories in four dimensions, the component  
reduction was originally carried out using the Wess-Zumino iterative procedure 
\cite{WZ} (see \cite{WB,GGRS} for reviews)
and its generalizations \cite{LR,Muller82,Ramirez,Muller,Muller89}.
Broadly speaking, this is a technique to reconstruct the relevant density multiplet from 
its lowest component and the known supersymmetry transformation law,  
in a suitably chosen Wess-Zumino gauge, 
order by order in powers of so-called covariant $\Q$-variables (of mysterious origin).
Although this technique can always be applied, at least in principle,  in practice it is rather 
awkward and (unreasonably) laborious.
As a result,  for some time the issue of component reduction remained the weakest point 
of superspace formulations for supergravity.\footnote{For supergravity theories 
possessing prepotential formulations, there exists an alternative, quite systematic scheme 
for component reduction \cite{BK}. Unfortunately, such prepotential formulations are not available 
in many cases.} 

This situation has changed with the  observation \cite{AD} that the concept of superspace normal coordinates \cite{McA} can be fruitful for component reduction in supergravity, 
which has led to the development of more powerful methods 
\cite{GKS,KT-M} (see also \cite{Ts}).
The crucial property  of the normal coordinate approaches
to component reduction is their universality. 
They can be used efficiently for any supergravity theory formulated in superspace, 
and for any number of space-time dimensions. 
At the same time, these methods are ultimately related to the earlier 
Noether-like schemes of \cite{WZ,WB,Muller82,Ramirez,Muller,Muller89},
for the fermionic normal coordinates (which correspond to parallel transport 
around the bosonic body of curved superspace)
can be seen to coincide with the  covariant $\Q$-variables \cite{KT-M}. 

Over a decade ago, a new universal method\footnote{The mathematical 
construction underlying the method of   \cite{Gates,GGKS} 
happens to be a special case of the theory of integration over surfaces in 
supermanifolds developed in \cite{GKSchwarz,BS,Vor}, 
see also \cite{KhN,Kh} for related reviews.} 
for component reduction in 
supergravity \cite{Gates,GGKS} was proposed,  sometimes referred to as ``ectoplasm,''
which appears to be  more radical  than the normal coordinate approach.
It presents a superform construction of supergravity actions, 
and is based on the use of a closed super $d$-form in the case of $d$ space-time dimensions.
Conceptually, it is very simple and its key points can be described in just  two paragraphs as follows.

Consider a curved superspace $ \cM^{d|\d}$ with $d$ space-time 
and $\d$ fermionic dimensions, and  let $\cM^{d|\d}$ be parametrized by local coordinates 
$z^M =(x^{\hat m} , \q^{\hat \m})$, where ${\hat m} =1, \dots, d$ and ${\hat \m}=1, \dots, \d$.
The corresponding superspace geometry is described by covariant derivatives 
\bea
\cD_A =(\cD_{\hat a} , \cD_{\hat \a})
=E_A +\F_A~, \quad 
E_A : = E_A{}^M\, \pa_M ~, \quad 
\F_A := \F_A  {\bm \cdot} {\mathbb J} =E_A{}^M \F_M ~.
\label{cov-der}
\eea
Here $\mathbb J$ denotes 
the generators of the structure group
(with all indices of $\mathbb J$s  suppressed),  
$E_A $ is the inverse vielbein, and $\F ={\rm d} z^M \F_M =E^A \F_A$ the connection. 
As usual, the matrices defining the vielbein $E^A := {\rm d}z^M E_M{}^A$
and its inverse  $E_A $ 
are such that
$E_A{}^M E_M{}^B =\d_A{}^B$ and
$E_M{}^A E_A{}^N =\d_M{}^N$. 
The covariant derivatives obey the algebra 
\bea
[\cD_A , \cD_B \} = T_{AB}{}^C \cD_C + R_{AB}{\bm \cdot} {\mathbb J} ~, 
\label{torsion}
\eea
with $T_{AB}{}^C $ the torsion, and $R_{AB}$ the curvature of $\cM$.

Next, consider a super $d$-form 
\be
J= \frac{1}{d!} {\rm d} z^{M_d}  \wedge \dots \wedge {\rm d}z^{M_1} J_{M_1\dots M_d}=
\frac{1}{d!} E^{A_d} \wedge \dots \wedge E^{A_1}  J_{A_1 \dots A_d}
\ee
constrained to be  closed  
\be
{\rm d} J =0~\qquad \Longleftrightarrow \qquad
\cD_{[B}J_{A_1\cdots A_d \}}-\frac{d}{ 2}T_{[BA_1|}{}^C J_{C|A_2\cdots A_d\}}=0~.
\label{cohomology}
\ee
Then, the integral over space-time
\bea
S&= &\frac{1}{d!} \int\rd^dx\, \ve^{{\hat m}_1 \dots {\hat m}_d }J_{{\hat m}_1 \dots {\hat m}_d}
=  \frac{1}{d!}\int\rd^d x\, \ve^{ {\hat m}_1 \dots {\hat m}_d}E_{{\hat m}_d}{}^{A_d} \dots E_{{\hat m}_1}{}^{A_1}
J_{A_1 \dots A_d}
\label{ectoplasm}
\eea
possesses the following fundamental properties: (i) $S$ is independent of the Grassmann variables
$\q$'s; and (ii) $S$ is invariant under general coordinate transformations on $ \cM^{d|\d}$ 
and structure group transformations, and therefore 
\bea
S&= &
\frac{1}{d!}\int\rd^d x\, \ve^{{\hat m}_1 \dots {\hat m}_d}E_{{\hat m}_d}{}^{A_d} \dots E_{{\hat m}_1}{}^{A_1}
J_{A_1 \dots A_d} \Big|_{\q=0}~. 
\label{ectoplasm2}
\eea
In physically interesting cases, the superform $J$ has to obey some additional covariant constraints
imposed on its components $J_{A_1 \dots A_d}$. This is how the dependence of $J$
on the geometric fields in (\ref{cov-der}) and (\ref{torsion}) occurs.

As is clear from the above discussion, the ectoplasm method is very general, 
and its use for component reduction\footnote{In the context of component reduction, 
its most recent application
has been given in \cite{GT-M} where the density projection formula for 2D $\cN=4$ 
supergravity was determined.} 
is just one of many possible applications. 
It is actually a method for constructing supersymmetric invariants. 
In particular, the method has already been applied to study the structure of higher-order corrections 
in heterotic string theory \cite{BH}, as well as  for elucidating the structure of higher-loop counterterms  
in maximally supersymmetric Yang-Mills theories \cite{BHS}.  The last two works are in
accord with comments made at the end of the first work in \cite{Gates}, where it was even
conjectured that the ectoplasmic concept might find application outside of supersymmetric
theories.

Independently of concrete applications, the starting point 
of the ectoplasm method is always a closed super $d$-form $J$ given explicitly. To construct such 
a superform, one has to address the cohomology problem (\ref{cohomology})
that is non-trivial in general. For instance, if one somehow fixes a non-vanishing 
component of $J_{A_1 \dots A_d}$ of lowest mass dimension and then tries to restore 
the components of higher dimension by iteratively solving the cohomology equations 
(\ref{cohomology}), the resulting calculation can be argued to be equivalent to that one 
encounters when applying the normal coordinate method 
of \cite{KT-M} (which proves to be more powerful than the scheme presented in \cite{GKS}).
Therefore, in the context of component reduction, one does not gain much
if the ectoplasm method is implemented iteratively.   
However, the present paper is aimed to show that in conjunction with additional ideas 
this method becomes  the most efficient approach to component reduction in supergravity.
 
Given a $2n$-dimensional symplectic manifold, its volume $2n$-form $\O$  
is known to coincide, modulo a numerical factor, 
with $\o^{\wedge n} \equiv \o \wedge \dots \wedge \o$, where $\o$ is the symplectic two-form, 
${\rm d}\o =0$.
In this paper we will try to mimic this result in the case of four-dimensional 
supergravity theories. Specifically, for a given supergravity theory, we will try 
to engineer the corresponding four-form $J$ from the wedge-product 
of closed two-forms. It turns out that for this purpose it is sufficient to play with self-dual vectors 
multiplets (as defined, e.g., in \cite{Siegel} in the flat case)
if $\cN=1$ and $\cN=2$ supergravities are considered. 

This paper is organized as follows. In section 2 we illustrate our approach by 
providing a new {\it simplest/shortest} derivation 
of the closed four-form \cite{GGKS} which corresponds to the chiral action principle 
within the old minimal formulation for $\cN=1$ supergravity.  
Using the  idea described in section 2,  in section 3 we derive a closed four-form 
that generates  the component form of the chiral action principle 
in $\cN=2$ conformal supergravity. The latter result is then recast in the form of a complete 
density projector formula for a general $\cN=2$ locally supersymmetric action. 
A brief discussion of the results 
is given in section 4. The paper is concluded with two technical appendices
in which the superspace geometries for $\cN=1$ old minimal and 
$\cN=2$ conformal supergravities are reviewed in a concise form.
 
\section{Chiral action in N = 1 old minimal supergravity}
\setcounter{equation}{0}

The closed four-form, which corresponds to the chiral action principle 
within the old minimal formulation for $\cN=1$ supergravity, was constructed in   \cite{GGKS}. 
As an illustration of our procedure, in this section  we present
a new,  simplest derivation of this superform.
It is based on the use of a self-dual vector multiplet.
The latter is described by a complex closed two-form
\be
F= \hf E^B \wedge E^A F_{AB}~, \qquad {\rm d} F =0
\ee
which is characterized by the following components:
\begin{subequations} 
\bea
F_{\a\b}=0 ~, &&\qquad F_A{}^{\bd}=0~,~~~
\label{N1-complex-Vector-1}
\\
F_{a\b}=-(\s_a)_{\b \bd}\bar{W}^\bd~,&&
\qquad
F_{ab}=
-{\ri\over 2}(\ts_{ab})^{\ad\bd} \,\cDB_\ad\bar{W}_{\bd}
~. 
\label{N1-complex-Vector-3}
\eea
\end{subequations} 
Here the spinor field strength $\bar{W}_\ad$ is covariantly antichiral, 
\be
 \cD_\a\bar{W}_\ad=0~, 
 \ee
and obeys the Bianchi identity 
\be
\cDB_\ad\bar{W}^\ad=0~
\label{N1-complex-Vector-4}
\ee
which implies  that the vector multiplet is on-shell.
In other words, the explicit expression for the two-form is
\bea
F&=& -E^\b\wedge E^a(\s_a)_{\b \bd}\bar{W}^\bd
-\frac{\rm i}{4} E^b \wedge E^a(\ts_{ab})^{\ad\bd}\,\cDB_\ad\bar{W}_{\bd}
~.
\eea

It is an instructive exercise to check explicitly, using the (anti)commutation relations 
 for the covariant derivatives $\cD_A$ collected in Appendix A, 
 that the complex two-form $F$ defined above is
indeed closed, ${\rm d} F =0$. 
Alternatively, the latter property becomes obvious if one recalls
the structure of an {\it off-shell real} vector multiplet in curved superspace 
(see \cite{WB,GGRS,BK} for reviews).
Its field strength 
\be 
\cF ={\rm d} \cV= \hf E^B \wedge E^A \cF_{AB}~, \qquad \cV = E^A \cV_A~,
\ee
with $\cV$  the gauge field, is characterized by the following components:
\begin{subequations} 
\bea
&\cF_{\a\b}=\cF_{\a \bd}=\cF_{\ad\bd}=0~,~~~
\label{N1Vector-0-1}
\\
&\cF_{\a,\b\bd}=2\ve_{\a\b}\bar{\cW}_\bd~,~~~
\cF_{\ad,\b\bd}=2\ve_{\ad\bd}\cW_\b~,
\label{N1Vector-0-2}
\\
&
\cF_{\a\ad,\b\bd}=
\ri\ve_{\a\b}(\cDB_\ad\bar{\cW}_{\bd})
+\ri\ve_{\ad\bd}(\cD_\a \cW_\b)~,
\label{N1Vector-0-3}
\eea
\end{subequations} 
where the spinor field strength $\cW_\a$ and 
its conjugate ${\bar \cW}_\ad$ 
obey the Bianchi identities 
\bea
\cDB_\ad \cW_\a=\cD_\a\bar{\cW}_\ad=0~,\qquad
\cD^\a \cW_\a=\cDB_\ad\bar{\cW}^\ad~.
\eea
If the equation of motion for a free vector multiplet
is imposed, 
$\cD^\a \cW_\a=\cDB_\ad\bar{\cW}^\ad =0$, the two sectors of 
$\cF$ which involve the chiral $\cW_\a$ and antichiral ${\bar \cW}_\ad$ field strengths, 
respectively, become completely independent, modulo the reality condition. 
The self-dual vector multiplet is formally obtained by setting $\cW_\a =0$ while keeping 
the other field strength  ${\bar \cW}_\ad$ non-vanishing. 

Consider the closed four-form $J = F \wedge F$, 
\be
J= \frac{1}{24} E^D \wedge E^C \wedge E^B \wedge E^A J_{ABCD}
~, \qquad {\rm d} J =0~.
\label{J-def}
\ee
Using eqs. (\ref{N1-complex-Vector-1}), (\ref{N1-complex-Vector-3}) and the relations given in  Appendix A, 
one can represent  the non-vanishing components of $J$ as follows:
\begin{subequations} 
\bea
J_{ab\g\d}&=&-8{\rm i}(\s_{ab})_{\g\d}
{\bar \cL}_{\rm c}
\label{J-com1}
~,
\\
J_{abc\d}&=&{\rm i}\,\ve_{abcd}(\s^d)_{\d\ad}\cDB^\ad {\bar \cL}_{\rm c}~,
\label{J-com2}
\\
J_{abcd}&=&-\frac{1}{4} \ve_{abcd}\Big(
{\bar \cD}^2 - 12R \Big) {\bar \cL}_{\rm c}
~.
\label{J-com3}
\eea
\end{subequations} 
Here ${\bar \cL}_{\rm c}$ is a covariantly antichiral scalar superfield, 
\be
{ \cD}_\a {\bar \cL}_{\rm c} =0~, 
\label{J-anti}
\ee
which is expressed in terms of the vector multiplet strength as 
${\bar \cL}_{\rm c} = \frac{\rm i}{2} {\bar W}^2$.
This representation for $ {\bar \cL}_{\rm c}$ is, however, completely irrelevant 
in order to demonstrate the fact that the four-form $J$ with the non-vanishing components 
(\ref{J-com1})--(\ref{J-com3}) is closed, for eq. (\ref{J-anti}) suffices.
At this stage, the self-dual vector multiplet has completed its role and can be forgotten.

Using the closed four-form $J$ associated with an arbitrary covariantly antichiral scalar superfield
${\bar \cL}_{\rm c} $, one can construct a locally supersymmetric action in accordance with the general 
rule (\ref{ectoplasm2}). 
It only remains to define 
the component vierbein $e_m{}^a :=  E_m{}^a|_{\q=0}$ 
and its inverse $e_a{}^m$, such that 
\be
e_a{}^me_m{}^b=\d_a^b~,\qquad
e_m{}^ae_a{}^n=\d_m^n~, \qquad e:= \det (e_m{}^a)~,
\ee
as well as the gravitino $\Psi_m{}^\a:=2 E_m{}^\a|_{\q=0}$ 
and its tangent-space version
$\Psi_a{}^\a := e_a{}^m\Psi_m{}^\a$. 
Then, for the action we obtain
\bea
S_{\rm c}&=&-\int\rd^4x\, e\Big(
{1\over 4}\cDB^2
-3R
-{\ri\over 2}(\s^{d})_{\d\ad}\Psi_d{}^\d \cDB^\ad
+(\s^{ab})_{\g\d}\Psi_a{}^\g\Psi_b{}^\d 
\Big) {\bar \cL}_{\rm c} 
\Big|_{\q=0}
~.
\label{N=1compaction}
\eea
This agrees with the results given in \cite{WB,GGRS,BK}.

\section{Chiral action in N = 2 conformal supergravity}
\setcounter{equation}{0}

We now turn to constructing a closed four-form destined to  generate the chiral action principle 
in $\cN=2$ conformal supergravity. As shown in \cite{KLRT-M,KLRT-M2}, 
$\cN=2$ conformal supergravity can be described using the superspace geometry 
proposed by Grimm \cite{Grimm} which is more economical than the formulation 
given in \cite{Howe} (more precisely, the former is obtained from the latter by partially fixing the 
gauge freedom including the super-Weyl invariance).
Appendix B contains all information about the geometric formulation of \cite{Grimm},
which is relevant for this paper. A complete presentation can be found in \cite{KLRT-M}.

A self-dual $\cN=2$ vector multiplet in curved superspace
is described by a complex  two-form, 
$F= \hf E^B \wedge E^A F_{AB}$, with  the following components:
\begin{subequations} 
\bea
F_{A\,}{}^{\bd}_j&=&0~, \\
F_{\a}^i{}_\b^j&=&-2\ve_{\a\b}\ve^{ij}\bar{W}~,\qquad
F_{a}{}_{\b}^j=
{\ri\over 2}(\s_a)_\b{}^{\gd}\cDB_\gd^j\bar{W} ~,~~~\\
F_{ab}&=&
-{\frac18} \Big( (\ts_{ab})_{\ad\bd} 
\Big[ \cDB^{\ad k}\cDB^{\bd}_k - 4 \bar{Y}^{\ad\bd} \Big]
+4({\s}_{ab})_{\a\b}{W}^{\a\b}
\Big)\bar{W} ~.
\label{F_com3}
\eea
\end{subequations} 
Thus  the field strength $F_{AB}$ is generated by a single scalar superfield $\bar W$ 
which is  covariantly antichiral, 
\be
\cD_\a^i \bar{W}=0~,
\ee
and subject to the equation of motion
\be
\Big(\cDB_\gd^{(i}\cDB^{\gd j)}+4 \bar{S}^{ij}\Big)\bar{W} =0~.
\ee
Here the tensor superfields $W^{\a \b}$, $ {\bar Y}^{\ad \bd}$ and ${\bar S}^{ij}$ are 
components of the superspace torsion, see Appendix B.
It follows from (\ref{F_com3}) that $F_{ab}$ possesses both self-dual ($F_{\ad\bd}$) 
and anti-self-dual ($F_{\a\b}$) components. 
\be
F_{\ad\bd} = \frac{1}{8}  \Big[ \cDB_\ad^k \cDB_{\bd k} - 4 \bar{Y}_{\ad\bd} \Big]\bar{W}~, 
\qquad  F_{\a \b} = -\hf W_{\a \b} \bar{W}~.
\ee
However, the latter originates solely due to the curved superspace geometry.
Indeed, $F_{\a\b}$ is proportional to  the super-Weyl tensor $W_{\a\b}$, 
and hence it vanishes in the flat superspace limit.

The above relations imply that the two-form $F$ is closed. 
\be
{\rm d} F=0~.
\ee
To justify this claim, it is sufficient to consider an off-shell real $\cN=2$ vector multiplet
in curved superspace, $\cF= \hf E^B \wedge E^A \cF_{AB}$, 
which is described in detail in \cite{KLRT-M}. 
Its components are expressed in terms of a covariantly chiral superfield $\cW$, 
${\bar  \cD}^i_\ad \cW=0$, and its conjugate $\bar \cW$ which are related to each other 
by the Bianchi identity\footnote{Eq. (\ref{vectromul}) is a curved-superspace extension 
of the Bianchi identity given in \cite{GSW}.}
\bea
\Big(\cD^{\g(i}\cD_\g^{j)}+4S^{ij}\Big)\cW
&=&
\Big(\cDB_\gd^{(i}\cDB^{ j) \gd}+ 4\bar{S}^{ij}\Big)\bar{\cW}
~.
\label{vectromul}
\eea
On the mass shell, when the expressions in both sides of  (\ref{vectromul}) 
vanish, one can consistently switch off $\cW$ while keeping $\bar \cW$ non-vanishing.
This results in the self-dual $\cN=2$ vector multiplet introduced.

Now, consider the closed four-form $J = F \wedge F$,  
\be
J= 
\frac{1}{24} E^D \wedge E^C \wedge E^B \wedge E^A J_{ABCD}
~, \qquad {\rm d} J =0~.
\label{N=2_J}
\ee
It is an edifying calculation to verify that the non-vanishing components of $J$ 
can be represented as follows:
\begin{subequations}
\bea
J_{\a}^i{}_\b^j{}_\g^k{}_\d^l&=&-32{\rm i} \Big(\ve_{\a\b}\ve_{\g\d}\ve^{ij}\ve^{kl}
+\ve_{\a\g}\ve_{\d\b}\ve^{ik}\ve^{lj}
+\ve_{\a\d}\ve_{\b\g}\ve^{il}\ve^{jk}
\Big)
{\bar \cL}_{\rm c}~,
\label{bottom}
\\
J_a{\,}_\b^j{}_\g^k{}_\d^l&=&
-4 \Big( \ve_{\g\d}\ve^{kl}(\s_a)_\b{}^{\ad}\cDB_\ad^j
+ \ve_{\d\b}\ve^{lj}(\s_a)_\g{}^{\ad}\cDB_\ad^k
+ \ve_{\b\g}\ve^{jk}(\s_a)_\d{}^{\ad}\cDB_\ad^l
\Big)
{\bar \cL}_{\rm c}~,
\\
J_a{}_b{\,}_\g^k{}_\d^l&=&
{\rm i} \Big(
\ve_{\g\d}\ve^{kl}(\ts_{ab})_{\bd\gd}\cDB^{\bd\gd}
+2
( \s_{ab})_{\g\d}\cDB^{kl}
+ 16
(\s_{ab})_{\g\d}\bar{S}^{kl}
\non\\
&&
-8\ve_{\g\d}\ve^{kl}\big(({\s}_{ab})_{\a\b}{W}^{\a\b}
-({\ts}_{ab})_{\ad\bd} \bar{Y}^{\ad\bd}\big)\Big)
{\bar \cL}_{\rm c}~,
\eea
\bea
J_a{}_b{}_c{\,}_\d^l&=&
-{\rm i}\,
\ve_{abcd}\Big({1\over 6}(\s^d)_{\d\ad}\cDB^{\ad}_q\cDB^{lq}
+{5\over 3}(\s^d)_{\d\ad}\bar{S}^{lq}\cDB^{\ad}_{q}
-(\ts^d)^{\bd\a}{W}_{\a\d}\cDB_\bd^{l}
\non\\
&&
-(\s^d)_{\d\ad}\bar{Y}^{\ad\bd}\cDB_\bd^{l}
+{4\over 3}(\s^d)_{\d\ad}(\cDB^{\ad}_{q}\bar{S}^{lq}) \Big)
{\bar \cL}_{\rm c}~,
\\
J_a{}_b{}_c{}_d&=&
\ve_{abcd}\Big(
{1\over 96}\big( \cDB^{ij}\cDB_{ij}
-
\cDB^{\ad\bd}\cDB_{\ad\bd} \big)
+{2\over 3}\bar{S}^{ij}\cDB_{ij}
-{1 \over 3}\bar{Y}^{\gd\dd}\cDB_{\gd\dd}
+{2 \over 3}(\cDB^i_\bd\bar{S}_{ij})\cDB^{\bd j}
\non\\
&&
+{1 \over 6}(\cDB^{ij}\bar{S}_{ij})
+3 \bar{S}^{ij}\bar{S}_{ij}
- \big(\bar{Y}^{\ad\bd}\bar{Y}_{\ad\bd}
-{W}^{\a\b}{W}_{\a\b}\big)
\Big)
{\bar \cL}_{\rm c}
~.
\label{top}
\eea
\end{subequations}
Here the scalar ${\bar \cL}_{\rm c}$ is covariantly antichiral, 
\bea
{ \cD}_\a^i {\bar \cL}_{\rm c} =0~,
\label{N=2antichiral}
\eea  
and is related to the vector multiplet strength as
${\bar \cL}_{\rm c} = \frac{\rm i}{4} {\bar W}^2$.
The operators $\cDB_{ij}$ and $\cDB_{\ad\bd}$ in (\ref{top})
are defined as 
\bea
\cDB_{ij}:=\cDB_{\gd(i}\cDB_{j)}^\gd~, \qquad 
\cDB^{\ad\bd}:=\cDB^{(\ad}_k\cDB^{\bd)k}~.
\eea
They possess the following  useful identities:
\begin{subequations} 
\bea
\cDB^\ad_i\cDB^{\bd\gd}{U}&=&
\frac{2}{3} \Big(
\ve^{\ad(\bd}\cDB^{\gd) k}\cDB_{ik}
-8
\ve^{\ad(\bd}\bar{S}_{ij}\cDB^{\gd) j}
-8\ve^{\ad(\bd}\bar{Y}^{\gd)\dd}\cDB_{\dd i}
+6\bar{Y}^{(\ad\bd}\cDB^{\gd)}_i
\Big) {U}~,~~~~~~~~
\\
\cDB^{\ad\bd}\cDB_{\ad\bd}{U}
&=&
-\Big{(}
\cDB^{ij}\cDB_{ij}
-8\bar{S}^{ij}\cDB_{ij}
+8\bar{Y}^{\gd\dd}\cDB_{\gd\dd}
-16(\cDB_{\ad i}\bar{S}^{ij})\cDB^{\ad}_j
\Big{)}
{U}
~,
\label{cDB-4-spin-isospin}
\eea
\end{subequations} 
with $U$ a scalar superfield.

Now comes the crucial point of our analysis. 
Given an arbitrary covariantly antichiral scalar superfield ${\bar \cL}_{\rm c} $, 
eq. (\ref{N=2antichiral}),
one can check that the four-form $J$ with components (\ref{bottom})--(\ref{top}) is 
closed.\footnote{In the flat superspace limit, our four-form $J$ reduces to the so-called ``chiral'' four-form constructed
in \cite{BisSiegel} (this observation is not quite obvious to make, however, because of  the ingenious notation
adopted in \cite{BisSiegel}).}  
At this stage, therefore, we can completely forget about the on-shell
vector multiplet $F$ and the explicit realization for  ${\bar \cL}_{\rm c} $ in terms of $\bar W$ 
given.

Using the closed four-form $J$ 
constructed,  we generate a locally supersymmetric action in accordance with the general 
rule (\ref{ectoplasm2}). In complete analogy with the $\cN=1$ case, 
we define the component vierbein $e_m{}^a :=  E_m{}^a|_{\q=0}$ 
and  the gravitino $\Psi_m{}^\a_i:= 2E_m{}^\a_i|_{\q=0}$.  
Then, the resulting action is 
\bea
S_c&=&\int\rd^4x\, e\Bigg(
{1\over 96}\cDB^{ij}\cDB_{ij}
-{1\over 96}\cDB^{\ad\bd}\cDB_{\ad\bd}
+{2\over 3}\bar{S}^{ij}\cDB_{ij}
-{1\over 3}\bar{Y}^{\gd\dd}\cDB_{\gd\dd}
+{2\over 3}(\cDB^i_\bd\bar{S}_{ij})\cDB^{\bd j}
\non\\
&&~~~
+{1\over 6}(\cDB^{ij}\bar{S}_{ij})
+3\bar{S}^{ij}\bar{S}_{ij}
-\bar{Y}^{\ad\bd}\bar{Y}_{\ad\bd}
+{W}^{\a\b}{W}_{\a\b}
\non\\
&&~~~
-\frac{\rm i}{6} \Psi_d{}^\d_l(\s^d)_{\d\ad}\Big(
\frac{1}{4}
\cDB^{\ad}_q\cDB^{lq}
-\frac{1}{4}
\cDB_{\bd}^l\cDB^{\bd\ad}
+7\bar{S}^{lq}\cDB^{\ad}_{q}
-5\bar{Y}^{\ad\bd}\cDB_\bd^{l}
+4
(\cDB^{\ad}_{q}\bar{S}^{lq})
\Big)
\non\\
&&~~~
+{\ri\over 2}\Psi_d{}^\d_l(\ts^d)^{\ad\a}{W}_{\a\d}\cDB_\ad^{l}
\non\\
&&~~~
+\Psi_c{}^\g_k\Psi_d{}^\d_l \Big(
{1\over 4}(\s^{cd})_{\g\d}\cDB^{kl}
+2(\s^{cd})_{\g\d}\bar{S}^{kl}
-\ve_{\g\d}\ve^{kl}({\s}^{cd})_{\a\b}{W}^{\a\b}
\non\\
&&~~~~~~~~~~~~~~~~~
-{1\over 8}\ve_{\g\d}\ve^{kl}(\ts^{cd})_{\bd\gd}\cDB^{\bd\gd}
-\ve_{\g\d}\ve^{kl}({\ts}^{cd})_{\ad\bd} \bar{Y}^{\ad\bd}
\Big)
\non\\
&&~~~
+{1\over 4}\ve^{abcd}(\s_a)_{\b\ad}\Psi_b{}^\b_j\Psi_c{}^\g_k\Psi_d{}_\g^k\cDB^{\ad j}
+{\ri\over 4}\ve^{abcd}\Psi_a{}^\a_i\Psi_b{}_\a^i\Psi_c{}^\b_j\Psi_d{}_\b^j 
\Bigg){\bar \cL}_{\rm c} \Big|_{\q=0}
~.
\label{N=2action}
\eea
This component action was first  computed by M\"uller \cite{Muller89}
using different techniques. 
Its derivation using the ectoplasm approach is one of the main results of our paper.

The covariantly antichiral scalar superfield ${\bar \cL}_{\rm c}$ can be represented in terms of 
an unconstrained scalar superfield $\cL$ as follows \cite{Muller,KT-M}
\bea
{\bar \cL}_{\rm c} &=& \D \cL ~, \non \\
\D &=&\frac{1}{96} \Big((\cD^{ij}+16{S}^{ij})\cD_{ij}
-(\cD^{\a\b}-16{Y}^{\a\b})\cD_{\a\b} \Big)
\non\\
&=&\frac{1}{96} \Big(\cD_{ij}(\cD^{ij}+16{S}^{ij})
-\cD_{\a\b}(\cD^{\a\b}-16{Y}^{\a\b}) \Big)~,
\label{chiral-pr}
\eea
where we have defined
\bea
\cD_{\a\b}:=\cD_{(\a}^k\cD_{\b)k}~,\qquad
\cD_{ij}:=\cD^\g_{(i}\cD_{j) \g}~.
\eea
In the special case when $\cL$ is real, $\bar \cL = \cL$, eq. (\ref{N=2action}) 
constitutes the component of the general action 
\be
\int \rd^4 x \,{\rm d}^4\q{\rm d}^4{\bar \q}\,E\, \cL~, 
\qquad E = {\rm Ber}(E_M{}^A)~.
\ee
It can be brought to a manifestly real form by adding to the right-hand side of (\ref{N=2action}) 
its complex conjugate.

\section{Discussion} 
\setcounter{equation}{0}

The traditional approaches to the component reduction in four-dimensional 
$\cN=1$ supergravity are reviewed in the textbooks \cite{WB,GGRS,BK}. 
These approaches are known to be extremely laborious. Our derivation of the component action principle
(\ref{N=1compaction}) took only a few hours of  calculation, and its technical description requires 
half a page only.
This shows that the ectoplasm method becomes very efficient if the problem
of constructing a required closed super $d$-form (in the case of $d$ space-time dimensions) 
can be re-cast as that of engineering this superform from some closed superforms of lower rank.
This idea was successfully applied in section 3 to construct
the closed four-form (\ref{N=2_J})--(\ref{N=2antichiral})
in $\cN=2$ conformal supergravity, which is associated with an arbitrary 
covariantly antichiral scalar superfield ${\bar \cL}_{\rm c}$
and generates the locally supersymmetric action (\ref{N=2action}). 
The four-form  (\ref{N=2_J})--(\ref{N=2antichiral}) is a new original result 
derived for the first time in the present  paper. 
As to the $\cN=2$ chiral action (\ref{N=2action}), 
it was computed twenty years ago by M\"uller\footnote{It was the last paper on supergravity  written by 
Martin M\"uller.} \cite{Muller89} using a technique 
closely resembling the normal coordinate construction of \cite{KT-M}.
Our derivation of the action (\ref{N=2action}) is much more simpler as compared with 
the calculation in \cite{Muller89}.

With the component action (\ref{N=2action}) at our disposal, the projective-superspace 
formulation for $\cN=2$ matter-coupled supergravity given in  
\cite{KT-M,KLRT-M,KLRT-M2,K_2008} is completely developed. In particular, 
any $\cN=2$ supergravity-matter action can be readily reduced to components.\\

\noindent
{\bf Acknowledgements:}\\
The work of SJG and  GT-M is supported by the endowment of the John S.~Toll 
Professorship, the University of Maryland Center for String \& Particle Theory, 
and National Science Foundation Grant PHY-0354401.  GT-M is happy to thank 
the School of Physics at the University of  Western Australia for the kind hospitality 
and support during part of this work.

\appendix

\section{N=1 old minimal supergravity} 
\setcounter{equation}{0}

Here we collect the key relations used in this paper
concerning the superspace geometry within the old minimal formulation 
for $\cN=1$  supergravity, see \cite{BK} for more details.
Our notation and conventions correspond to \cite{BK}; they are similar to 
those used in \cite{WB} except for the normalization of the 
Lorentz generators, including a sign definition of  
the sigma-matrices $\s_{ab}$ and $\tilde{\s}_{ab}$.

The superspace geometry is described by covariant derivatives
\bea
\cD_A &=& (\cD_a , \cD_\a ,\cDB^\ad ) = E_A+ \O_A~, \quad
 \O_A = \hf\,\O_A{}^{bc} M_{bc}
= \O_A{}^{\b \g} M_{\b \g}
+\bar{\O}_A{}^{\bd \gd} {\bar M}_{\bd \gd} ~,~~~
\eea
with  $\O_A$ the Lorentz connection
and $M_{bc} \Leftrightarrow ( M_{\b\g}, {\bar M}_{\bd \gd})$
the Lorentz generators, 
\bea
&{[}M_{\b\g},\cD_{\a}{]}
=\ve_{\a(\b}\cD_{\g)}~,\qquad
{[}\bar{M}_{\bd\gd},\cD_{\a}{]}=0~.
\eea
The covariant derivatives obey the following algebra:
\begin{subequations} 
\bea
\{\cDB_\ad,\cDB_\bd\}&=&
4R\bar{M}_{\ad\bd}
~,
\\
\{\cD_\a,\cDB_\bd\}&=&-2\ri\cD_\a{}_\bd~,\\
{[}\cDB_\ad,\cD_{\b\bd}{]}&=&
-\ri\ve_{\ad\bd}(R\cD_\b+G_\b{}^\gd\cDB_\gd)-\ri(\cD_\b R)\bar{M}_{\ad\bd}
\non\\
&&
+\ri\ve_{\ad\bd}(\cDB^\gd G_\b{}^\dd)\bar{M}_{\gd\dd}
-2\ri\ve_{\ad\bd}W_\b{}^{\g\d} M_{\g\d}
~,
\eea
\end{subequations}
where the tensors  $R$, $G_a = {\bar G}_a$ and
$W_{\a \b \g} = W_{(\a \b\g)}$ satisfy the Bianchi identities
\be
\cDB_\ad R= \cDB_\ad W_{\a \b \g} = 0~, \quad
\cDB^\gd G_{\a \gd} = \cD_\a R~, \quad
\cD^\g W_{\a \b \g} = {\rm i} \,\cD_{(\a }{}^\gd G_{\b) \gd}~.
\ee

\section{N=2 conformal supergravity} 
\setcounter{equation}{0}

This appendix contains a summary of the superspace geometry corresponding to 
$\cN=2$ conformal supergravity, see \cite{KLRT-M} for more details.  
Consider a curved 4D $\cN=2$ superspace  $\cM^{4|8}$ parametrized by 
local bosonic ($x$) and fermionic ($\q, \bar \q$) 
coordinates  $z^{{M}}=(x^{m},\q^{\mu}_\imath,{\bar \q}_{\dot{\mu}}^\imath)$,
where $m=0,1,\cdots,3$, $\mu=1,2$, $\dot{\mu}=1,2$ and  $\imath=\1,\2$.
The Grassmann variables $\q^{\mu}_\imath $ and $\teb_{\dot{\mu}}^\imath$
are related to each other by complex conjugation: 
$\overline{\q^{\mu}_\imath}=\teb^{\dot{\mu}\imath}$. 
Following \cite{Grimm},
the structure group is chosen to be ${\rm SL}(2,{\mathbb C})\times {\rm SU}(2)$,
and the covariant derivative 
$\cD_{{A}} =(\cD_{{a}}, \cD_{{\a}}^i,\cDB^\ad_i)$
have the form 
\bea
\cD_{A}&=&
E_{A}+\O_{A}{}^{\b\g}\,M_{\b\g}
+{\bar \O}_{A}{}^{\bd\gd}\,\bar{M}_{\bd\gd}
+\Phi^{~\,kl}_{A}\,J_{kl}
~.
\label{CovDev}
\eea
Here 
$J_{kl}=J_{lk}$
are the generators of SU(2), 
and  $\Phi_{{A}}{}^{kl}(z)$   the corresponding connection. 
The action of the SU(2) generators on the covariant derivatives is defined as follows:
\bea
&{[}J_{kl},\cD_{\a}^i{]} =\,-\d^i{}_{(k}\cD_{ l) \a}~, \qquad
{[}J_{kl},\cDB^{\ad}_i{]}
=\,-\ve_{i(k}\cDB^\ad_{l)}~.~~
\eea
The covariant derivatives obey the (anti)commutation relations
\begin{subequations} 
\bea
\{\cDB^\ad_i,\cDB^\bd_j\}&=&
-4\bar{S}_{ij}\bar{M}^{\ad\bd}
-2\ve_{ij}\ve^{\ad\bd}\bar{Y}^{\gd\dd}\bar{M}_{\gd\dd}
-2\ve_{ij}\ve^{\ad\bd}{W}^{\g\d}M_{\g\d}
\non\\
&&
-2\ve_{ij}\ve^{\ad\bd}\bar{S}^{kl}J_{kl}
-4\bar{Y}^{\ad\bd}J_{ij}~,
\label{acr2} \\
\{\cD_\a^i,\cDB^\bd_j\}&=&
-2\ri\d^i_j(\s^c)_\a{}^\bd\cD_c
+4\d^{i}_{j}G^{\d\bd}M_{\a\d}
+4\d^{i}_{j}G_{\a\gd}\bar{M}^{\gd\bd}
+8 G_\a{}^\bd J^{i}{}_{j}~,
\\
{[}\cD_a,\cDB^\bd_j{]}&=&
-\ri(\s_a)_\a{}^{(\bd}G^{\a\gd)}
\cDB_{\gd j}
+{\frac{\ri}2}
\Big(({\ts}_a)^{\bd\g}\bar{S}_{jk}
-\ve_{jk}({\s}_a)_\a{}^{\bd}{W}^{\a\g}
-\ve_{jk}({\s}_a)^{\g}{}_\ad\bar{Y}^{\ad\bd}\Big)
\cD_\g^k
\non\\
&&
+{\frac{\ri}2}\Big((\s_a)_\d{}^{\bd}T_{cd}{}^{\d}_{ j}
+(\s_c)_\d{}^{\bd}T_{ad}{}^{\d}_{ j}
-(\s_d)_\d{}^{\bd}T_{ac}{}^{\d}_{ j}\Big)
M^{cd}
\non\\
&&
+{\frac{\ri}2}
\Big(-(\s_a)^{\g}{}_\gd\d_{j}^{(k}\cD_{\g}^{l)}\bar{Y}^{\bd\gd}
-(\s_a)_{\g}{}^{\bd}\d_{j}^{(k}\cD_{\d}^{ l)}{W}^{\g\d}
+{\frac12}(\s_a)_\a{}^\bd\cD^{\a}_{ j}\bar{S}^{kl}\Big)
J_{kl}~,~~~
\eea
\end{subequations}
where 
\bea
T_{ab}{}_\gd^k&=&-{\frac14}(\s_{ab})^{\a\b}\cDB_\gd^{ k}Y_{\a\b}
+{\frac14}(\ts_{ab})^{\ad\bd}\cDB_{\gd}^{ k}\bar{W}_{\ad\bd}
-{\frac16}(\ts_{ab})_{\gd\dd}\cDB^{\dd}_{l}S^{kl}~.
\eea
Here the real four-vector $G_{\a \ad} $ and the complex 
tensors $S^{ij}=S^{ji}$, $W_{\a\b}=W_{\b\a}$, 
$Y_{\a\b}= Y_{\b\a}$ 
obey the Bianchi identities:
\begin{subequations}
\bea
&\cDB^\ad_{k}\bar{S}^{kl}+\cDB_\gd^{l}\bar{Y}^{\gd\ad}=0~,
~~~
\cDB^{\ad}_{(i}\bar{S}_{jk)}=\cD^{\a}_{(i}\bar{S}_{jk)}=0~,~~~
\cDB^{(\ad}_{i}\bar{Y}^{\bd\gd)}=0~,~~~
\cDB^\ad_iW^{\b\g}=0~,~~~~~~~
\\
&\cDB^{\ad}_{i}G^{\g\bd}=
{1\over 4}\cD^{\g}_{i}\bar{Y}^{\ad\bd}
-{1\over 12}\ve^{\ad\bd}\cD^{\g l}\bar{S}_{il}
+{1\over 4}\ve^{\ad\bd}\cD_{\d i}{W}^{\g\d}~,\\
&\cD_\g^k\cD_{\d k}W^{\g\d}
-\cDB_{\gd k}\cDB_{\dd}^k\bar{W}^{\gd\dd}=
4{W}^{\a\b}Y_{\a\b}
-4\bar{W}_{\ad\bd}\bar{Y}^{\ad\bd}
~.
\eea
\end{subequations}


\end{document}